# DIRECT DETECTION OF GRAVITY WAVES FROM NEUTRON STARS

Redouane Al Fakir[1][2] & William G. Unruh[2]
[1] *The MISC Institute, Vancouver, BC, Canada*
and
[2] *Gravitation & Cosmology Group, Dept. of Physics & Astronomy*
*University of British Columbia, Vancouver, BC, Canada*

**In light of the discovery of the first-ever double pulsar system, PSR J0737-3039, we re-examine an earlier proposal to directly detect gravity waves from neutron stars. That 1993 gravity wave detection proposal hypothesised an extremely tight and almost exactly edge-on binary pulsar system within one kiloparsec of the Earth, almost identical to the now famous double pulsar system discovered in 2003. The gravity waves targeted here are waves from the individual neutron stars, not the much longer waves from the binary as a whole. Some confusion about this alternative approach to gravity wave detection has persisted in the literature. Some authors deemed the proposal too optimistic, arguing that the effect would have only a $1/b^2$ dependence ($b$ is the impact parameter), and no $1/b$ dependence. Here, we re-derive the effect in more detail, and confirm the initial estimate that the effect would indeed include a $1/b$ dependence. In fact, PSR J0737-3039 is such a perfect realisation of the configuration hoped for in the original proposal, that the resulting gravity wave signature in pulsar timing residuals may exceed the original estimate by four orders of magnitude. A precisely predictable, coherent modulation in pulse time-of-arrival measurements of $10^{-8}$ *sec/sec* is possible. A one-year intermittent experiment (15 minutes/day around double pulsar eclipses) on an instrument comparable to the SKA ($10^{-6}$ *sec* resolution for $1$ *sec* of integration time) could thus directly detect gravity waves from individual neutron stars. If neutron stars turn out to radiate much weaker gravity waves than hoped for here, then such an experiment would still provide previously inaccessible constraints on neutron star physics, and contribute significantly to the development of quantum field theory in strong gravity.**

There are two quantities that are of interest each on its own right: $\tau \equiv \delta(\Delta t)$, the gravity wave modulation of the photon time-of-flight, and $\dot{\tau} \equiv d\tau/dt$, the rate of change of that modulation. For added safety, let us calculate $\tau$ and $\dot{\tau}$ separately, and then verify that one is the other's derivative. Furthermore, to facilitate comparison with the critical literature [30], we shall then re-derive one of the leading terms from scratch, using the same Fourier transform method followed in [30]. The three calculations will agree that the gravity wave effect in question does indeed have a $1/b$ dependence, as claimed in the original proposal [20]. In the process, we shall find that the derivation in [30], which concluded to no $1/b$ dependence, is in fact also correct in general, but it does not extend to the case of prime interest in [20], where the pulsar is in tight orbit around the source of gravity waves.

Consider a source of gravitational waves such as a rotating neutron star, and consider that a companion pulsar is orbiting in a highly inclined plane relative to the line-of-sight (i.e., close to an edge-on position). Then, the electromagnetic rays from the pulsar will intersect the gravitational waves from the neutron star in a zone where these waves are at their strongest. The hope is that this relatively strong gravity wave imprint will eventually be delivered to the observer in the form of a coherent modulation in the pulsar time-of-arrival measurements (there is also an astrometric effect [21-23,35] and an interstellar scintillation effect [24], which we focus on elsewhere [35]).

$\dot{\tau}$, the rate of change in the pulsar time-of-arrival, can be extracted from the geodesic equation:

(1) $$\frac{d(\delta p_0)}{d\lambda} = \frac{1}{2} h_{\alpha\beta,0} p^\alpha p^\beta ,$$

where $h$ is the gravitational wave perturbation of the metric, $\lambda$ is an affine parameter of the photon trajectory, and the $p$'s are the photon momenta. Then, to linear order in the gravity wave perturbation,

(2) $$\dot{\tau} = \frac{\delta p_0}{p^0} = \frac{1}{2} \int_{pulsar}^{Earth} h_{\alpha\beta,0} \frac{p^\alpha p^\beta}{p^{0\,2}} dt .$$

Let us choose out axis so that the z-axis points from the gravity wave source to the Earth, and the x-axis points from the gravity wave source to the point of closest approach of the photon (where $x$ equals the impact parameter $b$.)

To first order in h, we can assume that the photon trajectory here is very nearly a straight line (of impact parameter $b$) that is parallel to the z-axis. We can then use

(3) $$x^1 \equiv x \approx b \ \& \ p^1 \approx 0 \ , \ x^2 \equiv y \approx 0 \ \& \ p^2 \approx 0, \ x^3 \equiv z \ \& \ p^3 \approx p^0 \approx cte .$$

where $h$ is the gravitational wave perturbation of the metric, $\lambda$ is an affine parameter of the photon trajectory, and the *p*'s are the photon momenta. Then, to linear order in the gravity wave perturbation,

$$(4) \quad \dot{\tau} = \frac{\delta p_0}{p^0} = \frac{1}{2} \int_{pulsar}^{Earth} h_{\alpha\beta,0} \frac{p^\alpha p^\beta}{(p^0)^2} dt.$$

To make contact with [30], let us write $h^{\alpha\beta}$ as a function of the quadrupole moment of the gravity wave source as

$$(5) \quad \overline{h}^{00} = 2G \partial_{ij} \left( \frac{D_{ij}(t-r)}{r} \right),$$

$$(6) \quad \overline{h}^{0i} = -2G \partial_j \left( \frac{\dot{D}_{ij}(t-r)}{r} \right),$$

$$(7) \quad \overline{h}^{ij} = 2G \frac{\ddot{D}_{ij}(t-r)}{r},$$

where *G* is the gravitational constant, and

$$(8) \quad \overline{h}^{\alpha\beta} \equiv h^{\alpha\beta} - \frac{1}{2} h \eta^{\alpha\beta}.$$

Here, $r \equiv (x^2 + y^2 + z^2)^{(1/2)} \approx (b^2 + z^2)^{(1/2)}$, $\eta^{\alpha\beta}$ is the flat background metric, with respect to which indexes are raised and lowered in this linear treatment, and $h \equiv \eta^{\alpha\beta} h_{\alpha\beta}$.

Using the rectilinear approximations **(3)** in the geodesic equation **(4),** we get

$$(9) \quad \dot{\tau} = \frac{1}{2} \int_{pulsar}^{Earth} (h_{00} + 2h_{03} + h_{33})_{,0} \, dz.$$

It is then straightforward to show that $\dot{\tau}$ connects to the trace-reversed metric perturbation in **(5-7)** through:

$$(10) \quad \dot{\tau} = \frac{1}{2} \int_{pulsar}^{Earth} (h^{00} - 2h^{03} + h^{33})_{,0} \, dz = \frac{1}{2} \int_{pulsar}^{Earth} (\overline{h}^{00} - 2\overline{h}^{03} + \overline{h}^{33})_{,0} \, dz.$$

To make contact with the original proposal in [20] this time, let us focus on one frequency component of the gravitational wave, characterized by angular frequency $\Omega$:

(11) $\quad D_{ij} \equiv H_{ij} \exp[i\Omega(t - r - t_{ph})],$

where the $H_{ij}$ are constant in space and time, and $t_{ph}$ is a temporal phase.

Substituting (5-7) and (11) in (10) produces, after some algebra,

(12) $\quad \dot{\tau} = \int_{pulsar}^{Earth} dz \left\{ \dfrac{G}{r^5} i\Omega e^{i\Omega(z-r-t_{ph})} \left[ H_{11}(2b^2 - z^2 - i\Omega r(z^2 - 2b^2) - b^2\Omega^2 r^2) \right. \right.$

$- H_{22}(1 + i\Omega r)r^2 + 2bH_{13}\left((1 + i\Omega r)(3z - i\Omega r^2) - z\Omega^2 r^2\right)$

$\left. \left. - H_{33}\left((1 + i\Omega r)(b^2 + 2z^2 + 2i\Omega z r^2) + \Omega^2(r^4 + z^4 + b^2 z^2)\right) \right] \right\}.$

Let us now switch to the dimensionless null variable

(13) $\quad s \equiv \Omega\left(\sqrt{b^2 + z^2} - z\right). \quad s \to 0$ when $z \to +\infty$ and $s \to +\infty$ when $z \to -\infty$.

It changes (12) into

(14) $\quad \dot{\tau} = \int_{pulsar}^{Earth} ds \left\{ e^{-i\Omega t_{ph}} \dfrac{2G}{(b^2\Omega^2 + s^2)^4} i\Omega^3 e^{-is} \right.$

$\left[ H_{11}\left(b^6\Omega^6(i + 2s) + b^4\Omega^4 s(2 - 9is + 4s^2) + b^2\Omega^2 s^3(-20 - 9is + 2s^2) + s^5(2 + is)\right) \right.$

$+ H_{22}\left(b^6\Omega^6 i + b^4\Omega^4 s(2 + 3is) + b^2\Omega^2 s^3(4 + 3is) + s^5(2 + is)\right)$

$- H_{13}\left(4b^5\Omega^5 s(2i + s) + 8b^3\Omega^3 s^2(3 - is + s^2) + 4b\Omega s^4(-6 - 4is + s^2)\right)$

$\left. \left. - H_{33}\left(2b^4\Omega^4 s(-2 + 4is + s^2) + 4b^2\Omega^2 s^3(4 + is + s^2) + 2s^5(-2 - 2is + s^2)\right) \right] \right\}.$

One can then obtain the answer in closed analytical form:

(15) $\qquad \dot{\tau} = e^{-i\Omega t_{ph}} \left[ F(s_{Earth}) - F(s_{Pulsar}) \right]$,

$$F(s) = \frac{2G}{(b^2\Omega^2 + s^2)^3} i\Omega^3 e^{-is} \left\{ \left( b^4\Omega^4(1+2is) + 2b^2\Omega^2 s^2(2+is) \right) H_{11} \right.$$

$$\left. - \left( b^4\Omega^4 + 2b^2\Omega^2 s^2 \right) H_{22} - 4ib\Omega s^2 \left( b^2\Omega^2 - 2i + s \right) H_{13} + s^2 \left( 2b^2\Omega^2(is-1) + s^2(2is+1) \right) H_{33} \right\}$$

where some of the terms were rearranged using the fact that $D_{ij}$ is traceless (see **(25-29)**). Several $1/b$ and $1/b^2$ terms are identifiable in (15), or more easily in the following expansion:

**(16)**

$$F(z) = G e^{i\Omega\left(z-\sqrt{b^2+z^2}\right)} \left[ H_{33} \left\{ \Omega^2 \left( \frac{z}{b^2+z^2} - \frac{1}{\sqrt{b^2+z^2}} \right) + \frac{i}{4}\Omega \left( -\frac{3z}{(b^2+z^2)^{3/2}} - \frac{1}{b^2+z^2} \right) \right\} \right.$$

$$H_{11} \left\{ \Omega^2 \left( -\frac{z}{b^2+z^2} - \frac{1}{\sqrt{b^2+z^2}} \right) + \frac{i}{4}\Omega \left( \frac{3z}{(b^2+z^2)^{3/2}} + \frac{1}{b^2+z^2} + \frac{4}{b^2} + \frac{4z}{b^2\sqrt{b^2+z^2}} \right) \right\}$$

$$H_{22} \left\{ \frac{i}{4}\Omega \left( \frac{-z}{(b^2+z^2)^{3/2}} + \frac{1}{b^2+z^2} - \frac{4}{b^2} - \frac{4z}{b^2\sqrt{b^2+z^2}} \right) \right\}$$

$$\left. H_{13} \left\{ b\Omega \left( \frac{1+\Omega z}{(b^2+z^2)^{3/2}} + \frac{\Omega}{b^2+z^2} \right) + i \left( -\frac{2z}{b(b^2+z^2)^{3/2}} - \frac{2}{b(b^2+z^2)} \right) \right\} \right]$$

Or, considering that $H_{11} + H_{22} + H_{33} = 0$,

**(17)**

$$F(z) = G e^{i\Omega\left(z-\sqrt{b^2+z^2}\right)} \left[ H_{11} \left\{ \Omega^2 \left( -\frac{2z}{b^2+z^2} \right) + \frac{i}{4}\Omega \left( \frac{6z}{(b^2+z^2)^{3/2}} + \frac{2}{b^2+z^2} + \frac{4}{b^2} + \frac{4z}{b^2\sqrt{b^2+z^2}} \right) \right\} \right.$$

$$+ H_{22} \left\{ \Omega^2 \left( \frac{-z}{b^2+z^2} + \frac{1}{\sqrt{b^2+z^2}} \right) + \frac{i}{4}\Omega \left( \frac{2z}{(b^2+z^2)^{3/2}} + \frac{2}{b^2+z^2} - \frac{4}{b^2} - \frac{4z}{b^2\sqrt{b^2+z^2}} \right) \right\}$$

$$\left. H_{13} \left\{ b\Omega \left( \frac{1+\Omega z}{(b^2+z^2)^{3/2}} + \frac{\Omega}{b^2+z^2} \right) + i \left( -\frac{2z}{b(b^2+z^2)^{3/2}} - \frac{2}{b(b^2+z^2)} \right) \right\} \right]$$

If we neglect the gravity wave amplitude at the Earth, i.e., if we consider that the observer is at $z \to +\infty$, then $s_{Earth} \to 0$ (see **(13)**) and

(17)     $F(s_{Earth}) = F(z \to +\infty) = F(s \to 0) = \frac{2G}{b^2} i\Omega \{H_{11} - H_{22}\}$ .

At the other asymptotic end of the z-axis, $z \to -\infty$, we have $s_{Pulsar} \to +\infty$, so **(15)** yields

(17)     $F(z \to -\infty) = 0$ .

Therefore, if one makes the approximation or assumption that the pulsar is infinitely far from the gravity wave source, one gets the asymptotic result

(18)     $\dot{\tau}(z_{Pulsar} \to -\infty) = -\frac{2G}{b^2} \frac{\partial}{\partial t_{ph}} \{D_{11}(t_{ph}) - D_{22}(t_{ph})\}$ ,

where

(19)     $D_{ij}(t_{ph}) \equiv H_{ij} \exp[-i\Omega t_{ph}]$.

In addition to $1/b^2$ terms like those in **(18)**, equation **(15)** contains not one, but several $1/b$ terms, as one can see in the interesting example $z_{Pulsar} \sim 0$, or $s_{Pulsar} \sim b\Omega$, which is a case invoked explicitly in [20]:

(20) $\quad \dot{\tau}(z_{Pulsar} \sim 0) = \dot{\tau}(s_{Pulsar} \sim b\Omega) = e^{-i\Omega t_{ph}} \frac{2G}{b^2} i\Omega \left\{ \left(1 - \frac{e^{-ib\Omega}}{2}\right)(H_{11} - H_{22}) + e^{-i\Omega b} H_{13} \right\}$

$\quad\quad\quad + e^{-i\Omega t_{ph}} \frac{G}{b} \Omega^2 e^{-ib\Omega} \{H_{11} - 2H_{13} + H_{33}\}.$

**(18-19)** is the result arrived at in [30]. It was concluded from its $1/b^2$ dependence that the estimates in the original proposal [20] (which had a $1/b$ dependence) were too optimistic, and that the effect would be definitely unobservable. We disagree with this conclusion for two main reasons.

First, **(15)** and **(20)** show that the effect does indeed have a $1/b$ dependence. In fact, the $1/b$ terms are killed off only if one imposes exactly the asymptotic condition $s_{Pulsar} \to +\infty$, i.e., $z_{Pulsar} \to -\infty$. But this is far from being the case in the proposal [20], where in contrast the pulsar is very close to the gravity wave source, and is in fact in tight orbit around it. In the case of a pulsar and a gravity-wave producing neutron star locked in a tight binary pulsar, $z_{Pulsar}$ is effectively constrained to always remain in the vicinity of zero, compared to the distance of the system to the Earth. In the case of PSR J0737-3039, for example, $z_{Pulsar}$ never strays beyond one light-second, which is about one orbital radius.

Second, the $1/b^2$ terms themselves can become comparable to $1/b$ terms if one can find alignments that are so extreme that b decreases to about one reduced gravity wavelength $1/\Omega$. One can see from (20) for example that the $\Omega/b^2$ terms become comparable to the $\Omega^2/b$ terms when $b\Omega \sim 1$.. This is the situation hoped for in [20], and found to hold in PSR J0737-3039 some ten years later: The slower of the two pulsars of that double pulsar system has a period of about 2.8 secs, which means that it radiates gravity waves mainly at 2.8 secs and 1.4 sec. On the other hand, it just happens apparently by pure coincidence that the distance between the two neutron stars is also about 2.8 secs. And furthermore the orbit happens to be almost exactly edge-on. The result is that our impact parameter $b$ is confined to values so small that $1/b^2$ are significant over a sizeable portion of the orbit. We analyse the situation specifically for the PSR J0737-3039 configuration in a separate work [36].

Besides $\dot{\tau}$, the time modulation rate-of-change, the time modulation itself, $\tau$, can be of interest on its own right. Let's evaluate it separately. Its derivative should then be $\dot{\tau}$ in **(15).**

Because of the rectilinear approximations **(3)**, we can extract $\tau$ directly from the photon line element

(21) $\quad (1 - h_{00}) dt^2 - 2 h_{0j} dt dx^j - (\eta_{ij} + h_{ij}) dx^i dx^j = 0,$

which yields

(22)  $(1 - h_{00}) dt^2 \approx (1 + 2h_{03} + h_{33)}) dz^2$.

To linear order in h, the modulation in the integrated time-of-flight is then

(23)  $\tau \equiv \delta(\Delta t) \approx \frac{1}{2} \int_{pulsar}^{Earth} (h_{00} + 2h_{03} + h_{33}) \, dz$.

Using (5-8), (11), and (13), we find

(24)  $\tau = e^{-i\Omega t_{ph}} \left[ J(s_{Earth}) - J(s_{Pulsar}) \right]$,

$$J(s) = \frac{2G}{(b^2\Omega^2 + s^2)^3} \Omega^2 \, e^{-is} \left\{ b^4\Omega^4\left((1+2is)H_{11} - H_{22}\right) - 4ib\Omega s^2\left((s-2i) + b^2\Omega^2\right)H_{13} \right.$$

$$\left. + 2b^2\Omega^2 s^2\left((is+2)H_{11} - H_{22} + (is-1)H_{33}\right) - s^4\left(H_{11} + H_{22} - 2(is+1)H_{33}\right) \right\},$$

which, compared with **(15)**, verifies that $\dot{\tau} = -i\Omega\tau = \partial\tau/\partial t_{ph}$. This simple relationship does not hold, of course, when the orbital movement of the pulsar is taken into account. Then, the impact parameter's time dependence introduces an additional modulation at the orbital frequency [36]. Furthermore, as it is familiar in other problems involving timing experiments in binary systems, several corrections need to be evaluated to determine how the various time delay effects that are involved should translate for the observer [37].

We would now like to make contact with the actual calculation performed in [30]. Because the Fourier method followed there is quite different from the direct derivations above and in [20], we would like to show how our $1/b$ dependence can also be calculated using the same Fourier method as in [30].

Calculating all the terms in **(15)** via an excursion into the frequency domain can be somewhat tedious, necessitating multiple integrations in the complex plane per term, especially in the interesting case where the pulsar is not at infinity. It will suffice here to first reproduce the $1/b^2$ result in [30], which is valid for the asymptotic limit $z_{Pulsar} \to -\infty$, and then show how one of the $1/b$ terms in **(15)**—for example the $H_{33}$ term—arises in non-asymptotic cases such as the binary pulsar configuration in [20].

Let us then go back to the geodesic equation **(4).** The $h_{\alpha\beta}$ in that expression are related to the trace-reversed energy-momentum tensor by

(25) $\quad (\nabla^2 - \partial_t^2) h_{\alpha\beta} = -16\pi G \bar{T}_{\alpha\beta} \equiv -16\pi G \left( T_{\alpha\beta} - \frac{1}{2} \eta_{\alpha\beta} T^\sigma{}_\sigma \right)$.

$h$ and $T$ can be decomposed into their Fourier components:

(26) $\quad h_{\alpha\beta}(X) = \int \frac{d^4 K}{(2\pi)^4} \hat{h}_{\alpha\beta}(K) e^{iK.X}, \quad \bar{T}_{\alpha\beta} = \int \frac{d^4 K}{(2\pi)^4} \hat{\bar{T}}_{\alpha\beta}(K) e^{iK.X}$ .

where $X \equiv (x^1, x^2, x^3, t)$ and $K \equiv (K^1, K^2, K^3, \omega)$ are the 4-vectors for position and momentum. **(25)** then becomes

(27) $\quad \hat{h}_{\alpha\beta}(K) = \frac{16\pi G}{(\vec{K}^2 - \omega^2 - i\varepsilon_1 \omega)} \hat{\bar{T}}_{\alpha\beta}(K)$ ,

which is to be evaluated at the limit $\varepsilon_1 \to 0$.

In the quadrupole approximation, the effect of the gravity wave source's spatial frequencies can be neglected in comparison with the temporal frequencies, and the space-space components of $\hat{T}$ obey

(28) $\quad \hat{T}_{ij}(K) \approx \int T_{ij}(\vec{X}) e^{-i\omega t} d^3 X \, dt = \hat{T}_{ij}(\omega)$.

These $\hat{T}_{ij}(\omega)$ are related to the $D_{ij}$ in (5-7) by

(29) $\quad \hat{T}_{ij}(\omega) = \frac{-\omega^2}{2} D_{ij}(\omega) = \frac{-\omega^2}{2} \int dt \, e^{i\omega t} D_{ij}(t)$.

Finally, the temporal components of $\hat{T}$ are obtained from the space-space components by energy-momentum conservation:

(30) $\quad \hat{T}^{0j} = \frac{K_i}{\omega} \hat{T}^{ij} \quad \text{and} \quad \hat{T}^{00} = \frac{K_i}{\omega} \hat{T}^{i0} = \frac{K_i K_j}{\omega^2} \hat{T}^{ij}$ .

The photon trajectory can be approximated by

(31) $\quad \vec{X} = \vec{X}_b + \lambda \vec{p}, \quad t = t_b + \lambda p^0$ ,

where $\vec{X}_b$ and $t_b$ correspond to the event of closest approach (to zeroth order in h). Also,

let $\varphi$ and $\theta$ be polar coordinates such that

(32)   $K^1 = k\sin\theta\cos\varphi$, $K^2 = k\sin\theta\sin\varphi$, $K^3 = k\cos\theta$.

Hence, the exponent in **(26)** can be written

(33)   $K.X = \vec{K}.\vec{X} - \omega t = bk\sin\theta\cos\varphi - \omega t_b + \lambda(K^3 - \omega)p^0$.

Putting it all together, (4) becomes

(34)   $\dot{\tau} = \dfrac{8\pi Gi}{(2\pi)^4}\int_{pulsar}^{Earth} dt \int d^3\vec{K}\, d\omega \dfrac{e^{i(\vec{K}.\vec{X}-\omega t)}}{(\vec{K}^2 - \omega^2 - i\varepsilon_1\omega)}$

$$\omega D_{ij}(\omega)\, \dfrac{\omega^2 p^i p^j - 2\omega K^j p^0 p^i + K^i K^j (p^0)^2}{(p^0)^2}\quad .$$

Finally, after applying the rectilinear approximation **(3)** to **(34)**, we find for the full expression of $\dot{\tau}$ in Fourier space,

**(35)**

$\dot{\tau} = \dfrac{8\pi Gi}{(2\pi)^4}\int_{pulsar}^{Earth} p^0 d\lambda \int_{-\infty}^{+\infty} d\omega \int_0^\infty dk \int_0^{2\pi} d\varphi \int_0^\pi d\theta \sin\theta\, \dfrac{\exp i[bk\sin\theta\cos\varphi - \omega t_b + \lambda(K^3 - \omega)p^0]}{(k^2 - \omega^2 - i\varepsilon_1\omega)}$

$k^2\Big(\omega(K^3 - \omega)\{(K^3 - \omega)D_{33} + 2(D_{13}K^1 + D_{32}K^2)\} + D_{11}K^{1^2} + D_{22}K^{2^2} + 2D_{12}K^1 K^2\Big).$

We can now see how the asymptotic condition $(z_{Pulsar} \to -\infty)$ imposed in [20] kills off $1/b$ terms and leaves only $1/b^2$ terms in $\dot{\tau}$: Under that asymptotic condition, the $\lambda$ integral in **(35)** is evaluated from $-\infty$ to $+\infty$, thus producing a delta function $\delta(K^3 - \omega)$. That function in turn eliminates the low-$K$ terms in **(35)**, leaving only $(D_{11}K^{1^2} + D_{22}K^{2^2} + 2D_{12}K^1 K^2)$, which produces the $1/b^2$ terms in **(18).**

In contrast, there is no such delta function in our non-asymptotic case, because the $\lambda$ integration in **(35)** does not start from $-\infty$. Hence, the low-$K$ terms in **(35)** persist and cause the $K$ integral to produce $1/b$ terms as claimed in [20] and found above in **(15, 20)**.

Let us see explicitly the steps that lead from **(35)** to the $1/b$ and $1/b^2$ terms in **(15, 20).** Here are the main steps for the $D_{33}$ component, for example.

STEP 1: the $\lambda$ integration yields

(36) $\qquad \int_{pulsar}^{Earth} p^0 d\lambda \exp i[\lambda(K^3 - \omega)p^0] = \dfrac{\exp i[\Lambda (K^3 - \omega)p^0]}{(K^3 - \omega)p^0},$

where $\Lambda \equiv \lambda(pulsar) \neq -\infty$. Then, the $D_{33}$ component becomes

(37) $\qquad \dot{\tau}_{33} = -\dfrac{8\pi G}{(2\pi)^4} \int_{-\infty}^{+\infty} d\omega\, \omega \exp[-i\omega(t_b + \Lambda p^0)] \int_0^{2\pi} d\varphi \int_0^{\pi} d\theta \sin\theta \int_0^{\infty} dk$

$$k^2 (k\cos\theta - \omega)\, D_{33}\, \dfrac{\exp[ik(b\sin\theta\cos\varphi + \Lambda p^0 \cos\theta)]}{(k^2 - \omega^2 - i\varepsilon_1 \omega)}.$$

STEP 2: the $k$ integration of the various $\dot{\tau}(D_{ij})$ yields $1/q$ and $1/q^2$ terms, where $q \equiv (b\sin\theta\cos\varphi + \Lambda p^0 \cos\theta)$. As we shall see below, these terms will produce $1/b$ and the $1/b^2$ terms in $\dot{\tau}$, respectively. Below, we show explicitly all the subsequent steps for two terms yielded by the $k$ integration of $\dot{\tau}(D_{33})$ in **(37)**:

(38) $\qquad \dfrac{i\omega}{q}$ and $-\dfrac{\cos\theta}{q^2}.$

This will illustrate how the terms obtained by the direct method above and in [20] can be reproduced by the Fourier method in [30].

STEP 3: the $\theta$ integration of **(37)** can be performed in two parts, corresponding to the two terms in **(38)**.

First, we find

(39) $\qquad I_1(b, \Lambda p^0) \equiv i\omega \int_0^{\pi} d\theta\, \dfrac{\sin\theta}{b\sin\theta\cos\varphi + \Lambda p^0 \cos\theta} =$

$i\omega\, \dfrac{\pi b\cos\varphi + \Lambda p^0 \{\log(\Lambda p^0) - \log(-\Lambda p^0)\}}{b^2 \cos^2\varphi + (\Lambda p^0)^2},$

which yields

(40) $\qquad I_1(b, \Lambda p^0 \neq 0) = i\omega\, \dfrac{\pi}{b\cos\varphi - i|\Lambda p^0|}$

and $I_1(b, \Lambda p^0 \to 0) = i\omega \dfrac{\pi b \cos\varphi}{b^2 \cos^2\varphi + (\Lambda p^0)^2}$ .

Note that the second integral in **(37)** needs to be kept in that limiting form, in view of the subsequent $\varphi$ integration.

Second, defining

**(41)**  $I_2(b, \Lambda p^0) \equiv \displaystyle\int_0^\pi d\theta \, \dfrac{-\sin\theta \cos\theta}{(b\sin\theta\cos\varphi + \Lambda p^0 \cos\theta)^2}$ ,

we find

**(42)**  $I_2(b, \Lambda p^0 > 0 \text{ or } \Lambda p^0 \to 0) = \dfrac{-2\pi \Lambda p^0 b \cos\varphi + i\pi\left(b^2 \cos^2\varphi - (\Lambda p^0)^2\right)}{\left(b^2 \cos^2\varphi + (\Lambda p^0)^2\right)^2}$ ,

$I_2(b, \Lambda p^0 < 0) = \dfrac{-2\pi \Lambda p^0 b \cos\varphi - i\pi\left(b^2 \cos^2\varphi - (\Lambda p^0)^2\right)}{\left(b^2 \cos^2\varphi + (\Lambda p^0)^2\right)^2}$ .

STEP 4: the $\varphi$ integration of $I_1(b, \Lambda p^0)$ yields the same result for all values of $\Lambda p^0$, namely

**(43)**  $\displaystyle\int_0^{2\pi} d\varphi \, I_1(b, \Lambda p^0) = \dfrac{-2\omega\pi^2}{\sqrt{b^2 + (\Lambda p^0)^2}}$ ,

The same is true for $I_2(b, \Lambda p^0)$, for which we find

**(44)**  $\displaystyle\int_0^{2\pi} d\varphi \, I_2(b, \Lambda p^0) = \dfrac{-2\omega\pi^2 \Lambda p^0}{\left(b^2 + (\Lambda p^0)^2\right)^{3/2}}$ ,

STEP 5: the $\omega$ integration of **(43)** yields the following component of $\dot{\tau}$,

**(45)**  $\dot{\tau}_{33/1} = \dfrac{G}{2\pi\sqrt{b^2 + (\Lambda p^0)^2}} \displaystyle\int_{-\infty}^{+\infty} d\omega \, \omega^2 D_{33}(\omega) \exp[-i\omega(t_b + \Lambda p^0)]$ ,

where, as we recall from **(29)**,

(46) $\quad D_{33}(\omega) = \int_{-\infty}^{+\infty} dt \, e^{i\omega t} D_{33}(t)$.

The integration in **(45)** produces delta-function derivatives:

(47) $\quad \dot{\tau}_{33/1} = \dfrac{-G}{\sqrt{b^2 + (\Lambda p^0)^2}} \int_{-\infty}^{+\infty} dt \, D_{33}(t) \, \delta''[t - (t_b + \Lambda p^0)]$ .

Using the general property

(48) $\quad \int f(x) \, \delta^{(n)}[x] \, dx = -\int \dfrac{\partial f}{\partial x}(x) \, \delta^{(n-1)}[x] \, dx$ ,

we finally obtain

(49) $\quad \dot{\tau}_{33/1} = \dfrac{-G}{\sqrt{b^2 + z_{pulsar}^2}} \dfrac{\partial^2 D_{33}}{\partial t^2}(z_{pulsar})$ ,

where we have used **(1)**, **(31)** and $\Lambda \equiv \lambda(pulsar)$ to substitute $z_{pulsar} = t_{pulsar} = t_b + \Lambda p^0$.

Similarly, applying **(48)** to the omega integral of **(44)**, we find

(50) $\quad \dot{\tau}_{33/2} = \dfrac{-G \, z_{pulsar}}{(b^2 + z_{pulsar}^2)^{3/2}} \dfrac{\partial D_{33}}{\partial t}(z_{pulsar})$ .

As advertised, this shows explicitly that the terms (including $1/b$ terms) found by the direct integration method further above and in [20], can also be derived using the Fourier method followed in [30].

This concludes our proof that the claim in [30] that all $1/b$ terms are suppressed in the case of individual (or localised) sources of gravitational waves, does not apply to scenarios such as the one proposed in [20], namely when the pulsar is in the vicinity of the gravity wave source. Nevertheless, as we also confirmed above, the calculation in [30] is indeed correct when the pulsar is far removed from the gravity wave source, and so is the conclusion in [30] that earlier scenarios for the detection of individual gravity wave sources using pulsar timing [28] were ruled out.

CASE 1-The electromagnetic source and the gravity wave source are unrelated (see fig.1): In that case, and especially if the electromagnetic source is a pulsar, only a strike of extraordinary luck would bring the impact parameter *b* to the extremely small values needed for the effect to become observable. This is the scenario contemplated in earlier attempts at exploiting pulsar time delays for the detection of individual gravity wave sources [28].

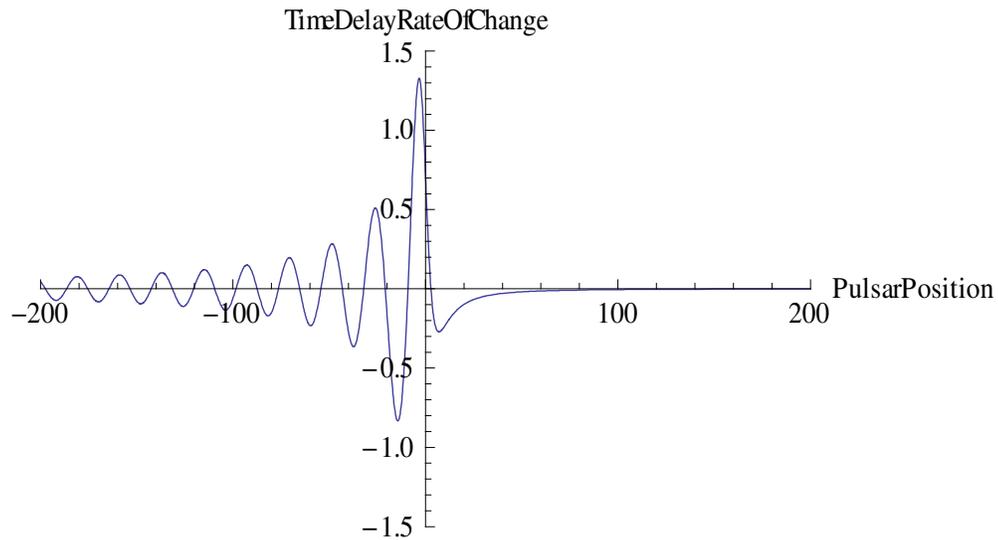

*FIGURE 1: The dimensionless time delay effect (in sec/sec) as a function of the pulsar position along the line-of-sight of the gravity wave source. The integrated effect (in sec) has the same behaviour, see equation (24). The pulsar position is measured in multiples of the gravitation wavelength. The time delay scale is arbitrary, the actual scale depending on how small the impact parameter b happens to be for a particular pulsar-gravity wave source combination. The figure, derived from equation (17), illustrates the behaviour of the time delay modulation depending on whether the pulsar is in the background (negative positions) or the foreground (positive positions) of the gravity wave source. In the latter case, the effect of the gravity waves on the electromagnetic pulses is strongly suppressed because then the pulses travel along with the waves instead of crossing through them.*

CASE 2-The electromagnetic source and the gravity wave source are part of the same gravitationally bound system (see fig.2): In this case, the extremely small impact parameters required to observe the effect might obtain automatically. This is the scenario described in [20], where a specific prescription is proposed in terms of a hypothetical binary pulsar where one of the neutron stars plays the role of gravity wave source and the other the role of electromagnetic source. The wish list for an ideal system specified that the binary should (1) be within about 1 kiloparsec of the Earth, (2) be almost exactly edge-on as viewed from the Earth, (3) include at least one fast pulsar, (4) have a very small orbital period. Of course, it is exactly such a system that was discovered some ten years later [29].

We study the particular case of the binary pulsar PSR J0737-3039 in more detail in [36], where the exact configuration in space of the system is taken into account. But we can already obtain an order of magnitude for the effect by taking the orbit to be circular and exactly edge-on. The big unknown here is of course the strength of gravity waves produced by an individual neutron star.

Since even the equation of state of such stars (and which parts are solid or liquid) is still unknown at this time, one can only speculate about the possible range of gravity wave amplitudes produced. Some amplitude estimates based on overall distortions of a rotating neutron star were obtained in the literature under various assumptions (see e.g. [33-35]).

Using these results as a guide for our ballpark estimates [20], we can consider that the quadrupole moment satisfies the relation $D_{ij}\Omega^2 \sim 10^{-2} \, cm$. Then **(15)** yields $\dot{\tau}(b \approx 1 \text{ lightsec}) \sim 10^{-12}$ as a typical gravity wave amplitude over most of the orbit [36]. That value climbs to $\dot{\tau}(eclipse) > 10^{-8}$ near the eclipse of the slower pulsar by the (much more intensely radiating) faster pulsar (see figure 2). The eclipse is determined by the size of the magnetosphere of the neutron star generating the gravity waves, which is about $10^4 \, km$.

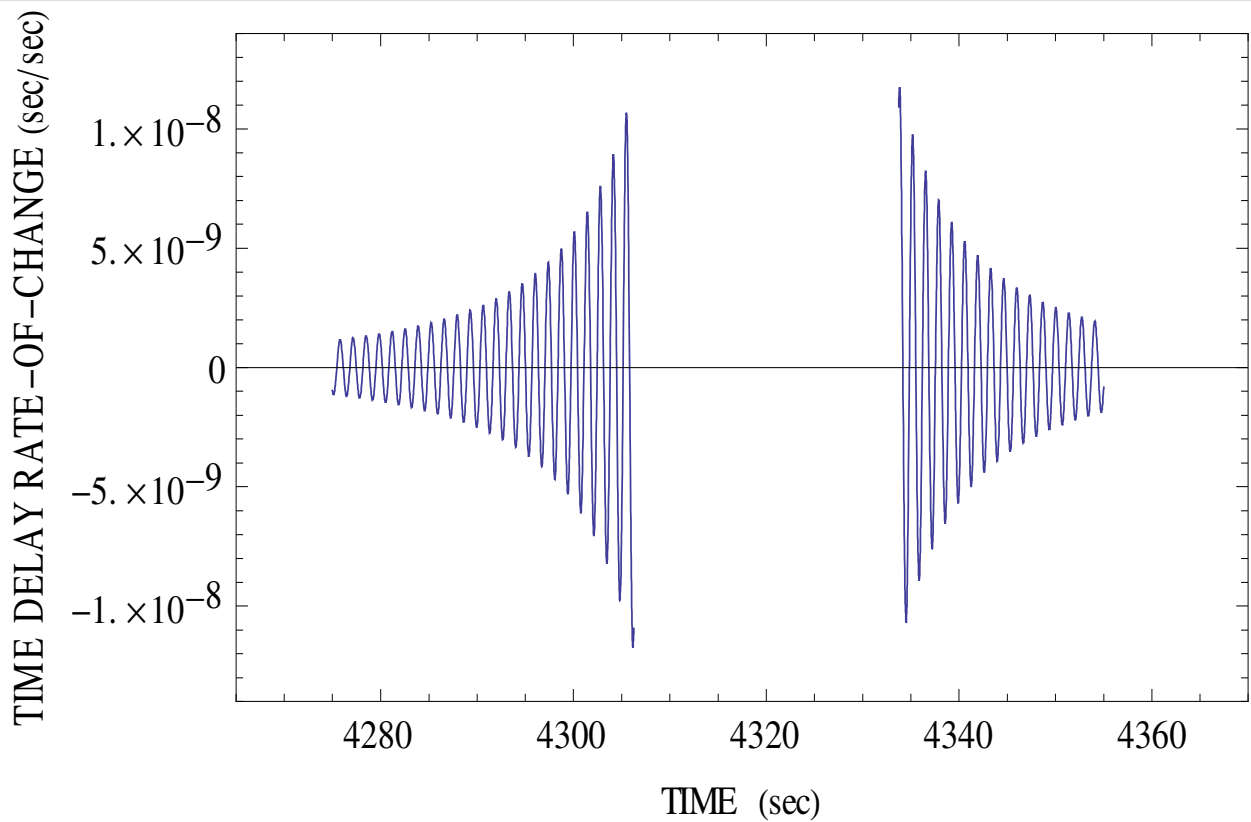

FIGURE 2: *The dimensionless time delay effect (in sec/sec) exerted by gravitational waves from the faster of the two pulsars of PSR J0737-3039 on the arrival times of pulses from the slower pulsar. For much of the orbit, the effect is at about $\sim 10^{-12}$ sec/sec. Around the eclipse by the magnetosphere of the faster pulsar, the effect can climb above $10^{-8}$ sec/sec. This quasi-ecliptic phase would last for about 30 secs every orbit period, that is every 2.4 hours. Since the slower pulsar has a period of about 2.7 secs, the total number of its pulses that cross the gravity waves of the faster pulsar during the quasi-ecliptic phase is more than $10^4$. Assuming a $\sqrt{N}$ statistical rule for the improvement in time resolution with the number of measurement points, this means that the effect could conceivably be detected by next generation radiotelescopes such as the SKA, which should be capable of measuring individual pulses with a $10^{-6}$ sec resolution.*

Acknowledgements

We are grateful to Sharon Morsink for several discussions on neutron stars. We also thank David Burg for helping to verify some of the calculations.